# On a Factorial Knowledge Architecture for Data Science-powered Software Engineering

Zheng Li
Department of Computer Science
University of Concepción
Concepción, Chile
imlizheng@gmail.com

## ABSTRACT

Given the data-intensive and collaborative trend in science, the software engineering community also pays increasing attention to obtaining valuable and useful insights from data repositories. Nevertheless, applying data science to software engineering (e.g., mining software repositories) can be blindfold and meaningless, if lacking a suitable knowledge architecture (KA). By observing that software engineering practices are generally recorded through a set of factors (e.g., programmer capacity, different environmental conditions, etc.) involved in various software project aspects, we propose a factor-based hierarchical KA of software engineering to help maximize the value of software repositories and inspire future software data-driven studies. In particular, it is the organized factors and their relationships that help guide software engineering knowledge mining, while the mined knowledge will in turn be indexed/managed through the relevant factors and their interactions. This paper explains our idea about the factorial KA and concisely demonstrates a KA component, i.e. the early-version KA of software product engineering. Once fully scoped, this proposed KA will supplement the well-known SWEBOK in terms of both the factor-centric knowledge management and the coverage/implication of potential software engineering knowledge.

## 1 Introduction

Since science is increasingly becoming data intensive and collaborative [1], data sharing gradually plays a crucial and beneficial role in various domains ranging from data-driven decision making to evidence-based software engineering [2]. As such, a fashionable trend is to make relevant data electronically available through central data repositories, in order to facilitate future and collaborative research and practices. In software engineering particularly, there have been international efforts on depositing software data including source code, bug history, benchmarking results, program logs, etc., in order to foster data-driven studies on software analysis, evolution and reengineering by different researchers and practitioners at different times.

Nevertheless, the immediately available data in repositories are not necessarily reusable directly. In fact, without suitable understanding and interpreting, the lakes of data can quickly turn into data swamps [3]. Take the field of software effort estimation as an example, if the relevancy is not carefully filtered, using imported data will bring locality-specific biases and deliver extremely poor estimation results [4]. Furthermore, by including different factors and/or different factor values, various datasets could even be incompatible with each other, not to mention their relevancy. For instance, there are 9, 18, and 26 factors in the Desharnais[1], COCOMO81[2], and Maxwell[3] datasets respectively; and when it comes to the factor *Application Type*, COCOMO81 distinguishes between three software project types, while Maxwell considers five application types [5]. In addition, considering the evolution of software technologies and methodologies, the old project data could have been far out of date to fit in today's software industry. For example, compared to a decade ago, the current cloud and microservice-based applications are generally developed and delivered on a continuous basis.

Meanwhile, although the discipline of data science is booming and promising for obtaining insightful and actionable knowledge from the deposited data (e.g., mining software repositories), the mining activities like pattern identification and prediction could be meaningless and even impossible "without knowing what to look for" [3]. More importantly, due to the lack of a "knowledge repository", the results of software data-driven studies are largely distributed and poorly organized, and consequently they are hard to be systematically shared and reused.

---

[1] http://promise.site.uottawa.ca/SERepository/datasets/desharnais.arff
[2] http://promise.site.uottawa.ca/SERepository/datasets/cocomo81.arff
[3] https://zenodo.org/record/268461#.XerFVehKjIU

Therefore, starting with a suitable knowledge architecture (KA) of software engineering tends to be a prerequisite of revealing insights from those software data lakes [3]. By employing a general system model [6] as a scaffold, we propose to develop a factorial KA from the existing software data repositories, in order to facilitate mining, storing, utilizing and sharing software engineering knowledge. For the purpose of conciseness, this paper only demonstrates our early achievement through the relatively well-established component KA of software product engineering.

This work makes two main contributions to the software engineering community. Firstly, our KA can act as a factor-oriented knowledge landscape for data science-powered software engineering based on the existing repositories (e.g., PROMISE[4] and FLOSSmole[5]), as emphasized in [3]. Note that the object in our factorial KA is software engineering knowledge instead of the huge amount of software data (see Section 2.1), and it is the organized factors and their relationships that essentially help guide/inspire future software data-driven studies. Secondly, the developed KA will be able to supplement the body of knowledge for software engineering (SWEBOK) [7], [8]. SWEBOK's intent is to include "generally accepted knowledge", whereas a KA offers spaces for both the existing and the future knowledge [9]. As such, our factorial KA will be extendable and supportive for diverse analyses of software data, rather than constraining them within a predefined schema.

The remainder of this paper is organized as follows. Section 2 clarifies the essential concepts of, and the relationships between, data, information and knowledge. In particular, Section 2.2 briefly explains KA in a generic sense, while Section 3 specifies our development of the factorial KA for data science-based software engineering. Conclusions and some future work are discussed in Section 4.

## 2 Essential Concepts

### 2.1 Data, Information, and Knowledge

In essence, knowledge and data are inextricably interwoven with each other through information. According to the clarification of relevant concepts [10], data are observation products with their representation symbols. Pure data are generally of little value due to the lack of context. Therefore, to make sense of data, the context (e.g., the structures and/or relationships of data) is needed to organize data into information. In specific, information can be defined as a function of data [10], for containing both the data and their context, as specified in Equation (1).

$$Information = f(Data) = Data + Context_d \quad (1)$$

where $f(Data)$ represents the function that makes sense of $Data$ and returns $Information$, and $Context_d$ indicates the context of $Data$.

Compared to information that is content-oriented and objective descriptions, knowledge is considered to be people-oriented and subjective interpretations [10]. Since accurate and adequate information plays a crucial and fundamental role in knowledge-intensive work [11], knowledge is also defined as the process of making sense of information. In addition, it has been identified that the context of information is essential for tailoring the interpretations to appropriate knowledge, and thus knowledge needs always to carry on the relevant context [10], [12]. Moreover, considering that the term "information" is used to indicate explicit descriptions, we further use "insight" to represent the tacit implications behind information. In this way, we do not have to distinguish between tacit and explicit knowledge by recognizing how the base information is gained and interpreted. Accordingly, we specify the definition of knowledge into Equation (2).

$$Knowledge = p(Information) \\ = Information + Context_i + Insight \quad (2)$$

where $p(Information)$ denotes the processing function that returns $Knowledge$ by making sense of $Information$ under its context $Context_i$.

### 2.2 Knowledge Architecture (KA)

Emerging from knowledge modelling and knowledge representation [11], KA is a discipline of creating, storing, sharing and utilizing human knowledge from the organizational perspective.

NASA defines KA as a combination of information architecture, knowledge management, and data science [13], as specified in Equation (3).

$$Knowledge\ Architecture = Data\ Science + \\ Knowledge\ Management + \\ Information\ Architecture \quad (3)$$

In detail, based on the structured and "shared information environments with useful, navigable form that resists entropy and reduces confusion" [14], KAs involve both knowledge extraction and knowledge managerial processes [15] by focusing on the building blocks of knowledge for specific applications [11].

Unlike information architecture that aims at catering the existing and known information entities, KA is supposed to deal with not only the existing but also the potential and future knowledge assets [9]. In particular, Data Science acts as the essential approach to transforming data to knowledge when establishing knowledge architectures [13].

There are various concerns about, and different approaches to, developing KAs. For example, a generalizable development strategy is claimed to cover knowledge characteristics, dimensions and resources [16]. Our factorial KA for data science-powered software engineering mainly focuses on the factors related to software projects, as specified in the next section.

## 3 Factorial KA of Software Engineering

---

[4] http://promise.site.uottawa.ca/SERepository/

[5] https://flossmole.org/

Recall that KA is generally implemented within individual organizations, and also recall that software systems could mirror the organizational structure they are designed in or designed for (i.e. Conway's Law [17]). By analogy, we treat a software project as a generic organizational concept, and thus the relevant processes and activities involved in the software project can be viewed as organizational behaviors. As such, we are convinced to argue the existence of a KA from the software project's perspective.

When it comes to a software project, the consensus is that different aspects and stages of the project are influenced by numerous and various factors. For instance, researchers have been concerned with both risk and success factors for software projects [18], [19], practitioners have tried to identify dominant factors from the existing applications and then utilize them to guide building new systems/services [20], while many models about software development (e.g., COCOMO) are proposed directly using relevant factors (e.g., cost drivers). Considering that factors play a vital role in driving the practices and research in the software engineering domain, we decided to develop a factorial KA of software engineering.

In fact, by screening the existing software repositories, it can be seen that most of the software data are collected around a set of factors. In other words, the software repositories are developed generally with factor-oriented schemas. Although the detailed project data could have been out of date, those various factors involved in software projects can always be valid. Thus, given a suitable structural space [9], [11], it will be feasible and valuable to distill the deposited time-sensitive data into a long-lasting factorial KA for data science-powered software engineering.

### 3.1 A System Model of Software Projects

Inspired by the factor-based system model [6], we also treat individual software projects as a general system by distinguishing between different types of factors, as illustrated in Figure 1. In detail,

- **Input Factors** play a prerequisite role in unfolding a software project. Since software projects are essentially driven by user requirements, the input factors are generally associated with requirements.
- **Inherent Factors** reflect software projects' properties that cannot be changed beyond certain thresholds. In a software project, this type of factors depend on the aforementioned input requirements only, while being independent of the project's environment.
- **(Controllable) Environmental Factors** affect the delivery process of a software product from its outside world, and the factors' effects can be controlled by adjusting their factor levels (or values) intentionally.
- **(Uncontrollable) Environmental Factors**, similarly, also influence the delivery process of a software product externally, while the factors' effects could be uncertain due to dynamic environmental situations.

- **Responses (Output Factors)** are observable variables of finished software projects. These output variables are essentially corresponding to the input factors, which meanwhile is under various influences from the other types of factors.

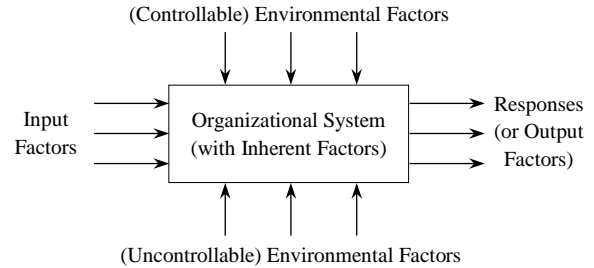

**Figure 1: A general system model of software projects from the factorial perspective.**

### 3.2 Fitting Factors into the Software Project Model

Due to the large amount of categories of datasets in the repositories (e.g., [5]), it is impossible to include the factors all at once in this paper. Therefore, we focus on the category of software effort estimation to arrange and fit factors into the software project model (see Figure 1). In specific, the factors are selected by referring to the software project datasets Albrecht[6], COCOMO, Desharnais, Maxwell, and the 14 general system characteristics in Function Point Analysis[7], while their arrangement is illustrated in Figure 2 and briefly explained as follows.

- Most requirement considerations are classified to be input factors. A particular characteristic of input factors is that their effects could be significantly influenced by environmental factors. For example, in a software project, the required development schedule might not be exactly satisfied due to mismatching personnel capacities, and the required system reliability will also depend on the system's hardware platform.
- Although closely related to requirements as well, the fixed product properties are regarded as a software project's inherent factors. For example, the application type will never be changed once it is defined, the involved data and functional features and counts are essentially part of the software product, while the corresponding product complexity has been identified to be neither reducible nor simplify-able [21].

---

[6] https://zenodo.org/record/268467#.XerOQehKjIU

[7] https://www.softwaremetrics.com/fpafund.htm

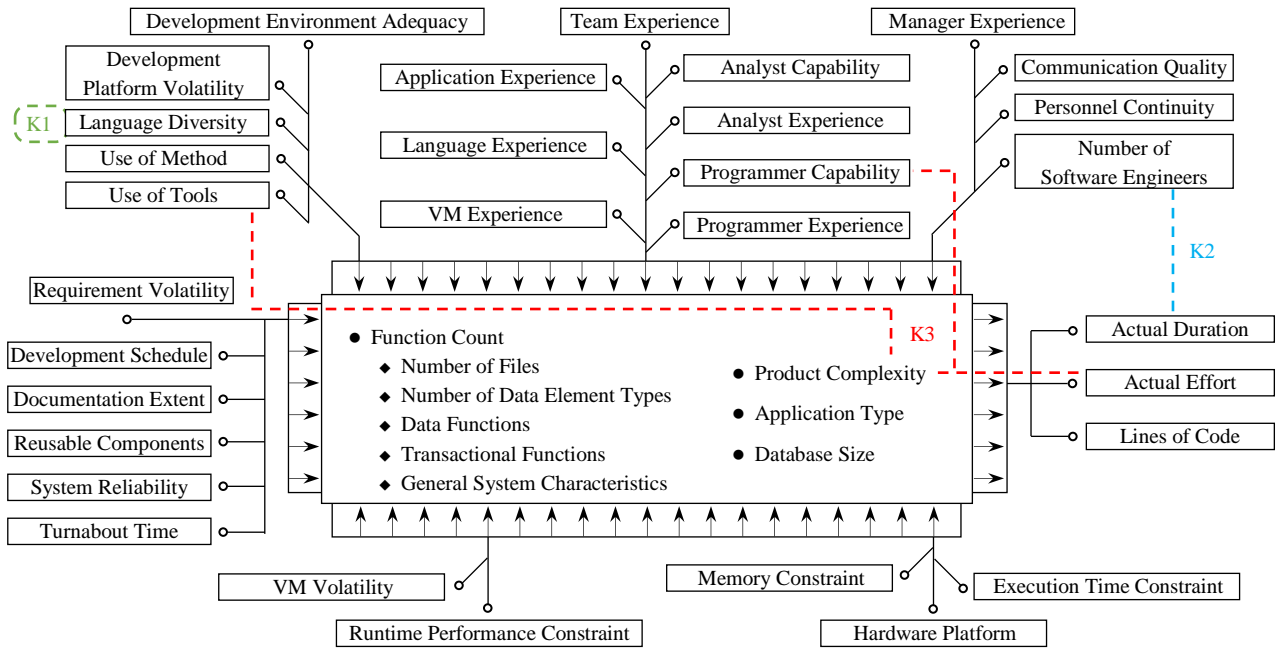

**Figure 2.** An example piece of knowledge architecture of software engineering. The knowledge can be managed by sticking to a single factor (e.g., K1 for Jones's Law), by correlating two factors (e.g., K2 for Brooks' Law), and by involving multiple factors (e.g., K3 for Li-O'Brien-Yang's Law).

- As explained in Section 3.1, things having outside impacts on the delivery of a software product can be viewed as environmental factors. From the perspective of software products, in particular, personnel attributes can also be treated as a special type of environmental factors. We consider environmental factors to be controllable if they take effects during the product development process. For example, it is possible to speed up a software project by improving the development environment adequacy and employing experienced development team and manager.
- In contrast, we consider environmental factors to be uncontrollable if they take effects at product runtime. For example, the program runtime performance could be unpredictable due to unknown hardware circumstances, and the virtual computing environment would particularly be unstable due to uncertain resource competitions.
- Given the investigated datasets, we highlight three observable variables as output factors of software projects, such as actual duration, actual effort, and lines of code (LOC). Note that, unlike function counts and database size, LOC is not a fixed product property because it could vary depending on programming language and programmer capacity. Therefore, we do not consider LOC to be an inherent factor of software projects.

### 3.3 Factorial Knowledge Architecture

Given the factors identified from the field of software effort estimation, what we demonstrate here is essentially a component KA of software product engineering [8]. However, the factors involved in this demonstration can further elicit other knowledge categories and areas in software engineering. For example, the factor language diversity is linked to the programming languages knowledge area of the computing fundamentals category, and the factor development schedule is associated with the software project management knowledge area of the software management category [7], [8].

Overall, such a KA can then facilitate hierarchically organizing, managing, discovering and sharing software engineering knowledge around those factors from the perspective of software projects. For the purpose of conciseness, we only highlight three scenarios of knowledge discovering/organizing as follows.

To begin with, software engineering knowledge can be formulated even by sticking to a single factor. Still take the factor language diversity as an example, the explicit information behind this factor can be those dozens of programming languages. Given the past decades' empirical data of software projects as information context, the statistics of programming language utility lead to one of Jones's Laws [22] (see K1 in Figure 2):

> *"In every decade, less than 10% of the available programming languages are used to code over 90% of all software applications created during that decade."*

Then, knowledge can be associated with the interactions between a pair of factors. For example, the COCOMO-based software engineering economics has revealed rich effort multiplier knowledge about how individual cost drivers causally link to actual effort in software projects. Besides the COCOMO-related knowledge, here we employ the well-known Brooks' Law (see K2 in Figure 2) as a demonstration):

> "Adding manpower to a late software project makes it later."

After examining the factor pairs, it is natural to focus on three-factor software engineering knowledge. For example, a data mining study shows that, to certain extents, a more complex project would inevitably employ highly capable programmers and use advanced tools, which will play a "friction" role in weakening the effect of product complexity on actual effort for software development, as described in Li-O'Brien-Yang's Law [21] (see K3 in Figure 2):

> "Increasing product complexity will result in interactions with other factors that could weaken and even overwhelmingly weaken the complexity's influence on actual effort in software projects."

Similarly, further knowledge of software engineering can be catered by gradually involving more factors and using data science techniques to identify their causal relationships.

## 4 Conclusions and Future Work

Driven by the data-intensive and collaborative trend in science, the software engineering domain also encourages uncovering useful and valuable insights through analyzing the shared data in software repositories. However, the revealed knowledge of software engineering is largely distributed and poorly organized due to a lack of "knowledge repository". To better utilize software repositories and enhance the value of relevant data-driven studies, we propose a factorial KA that uses software project factors and their relationships to facilitate data science-powered software engineering and to manage the distilled knowledge. It is noteworthy that such a KA is essentially to deal with the abstract knowledge instead of systematizing the huge amount of original data.

A possible threat to our work is that it could be too ambitious to develop the factorial KA of software engineering all at once. Thus, we plan to scope this KA on a field-by-field basis (e.g., the demonstrated field of software effort estimation in Section 3.3). Eventually, the overlapped factors and the linkages between different fields will naturally conjoin the KA components into a whole.

The developed factorial KA for data science-powered software engineering will be able to supplement (instead of replacing) the well-known SWEBOK. Firstly, unlike SWEBOK that divides software engineering knowledge into categories, areas and units, our KA provides a factor-centric approach to knowledge retrieval and management. Secondly, unlike SWEBOK that covers the mature knowledge only, our KA represents a K (number of factors) dimensional space that also includes the places for catering potential/future knowledge of software engineering.

## ACKNOWLEDGMENTS

This work is supported in part by Chilean National Commission for Scientific and Technological Research (CONICYT, Chile) under Grant FONDECYT Iniciación 11180905.

## REFERENCES

[1] Tenopir, C., Allard, S., Douglass, K., Aydinoglu, A. U., Wu, L., Read, E., Manoff, M., and Frame, M. 2011. Data sharing by scientists: Practices and perceptions. *PLoS ONE* 6, 6 (June 2011), e21101. DOI: https://doi.org/10.1371/journal.pone.0021101.

[2] Menzies, T., Kocaguneli, E., Turhan, B., Minku, L., and Peters F. 2014. Sharing Data and Models in Software Engineering, 1st ed. Morgan Kaufmann, Waltham, MA.

[3] Earley, S. March 2016. Want data insights? Start with your knowledge architecture. Retrieved August 9, 2020 from http://www.cmswire.com/big-data/want-data-insights-start-with-your-knowledge-architecture/

[4] Kocaguneli, E., Gay, G., Menzies, T., Yang, Y., and Keung, J. W. 2010. When to use data from other projects for effort estimation. In *Proceedings of the 25th IEEE/ACM International Conference on Automated Software Engineering*. (Antwerp, Belgium, September 20-24 2010). ASE '10. ACM Press, 321–324. DOI: https://doi.org/10.1145/1858996.1859061.

[5] Menzies, T., Krishna, R., and Pryor, D. 2016. The promise repository of empirical software engineering data. Department of Computer Science, North Carolina State University. Retrieved June 9, 2020 from http://promise.site.uottawa.ca/SERepository/

[6] Montgomery, D. C. 2019. *Design and Analysis of Experiments*, 9th ed. John Wiley & Sons, Inc., Hoboken, NJ.

[7] Bourque, P. and Fairley R.E.D. (Eds.) 2014. *Guide to the Software Engineering Body of Knowledge*, 3rd ed. IEEE Computer Society, Piscataway, NJ.

[8] Hilburn, T. B., Hirmanpour, I., Khajenoori, S., Turner, R., and Qasem, A. 1999. *A software engineering body of knowledge version 1.0*. Technical Report CMU/SEI-99-TR-004. Software Engineering Institute, Carnegie Mellon, Pittsburgh, PA.

[9] Gent, A. March 2008. What is knowledge architecture (the short version). Retrieved August 9, 2020 from https://incrediblydull.blogspot.jp/2008/03/what-is-knowledge-architecture-short.html

[10] Kaipa, P. 2000. Knowledge architecture for the twenty-first century. *Behaviour & Information Technology* 19, 3 (2000), 153–161. DOI: https://doi.org/10.1080/014492900406146.

[11] Sandkuhl, K. 2015. Pattern-based knowledge architecture for information logistics. *Revista Investigacion Operacional* 36, 1 (2015), 36–44.

[12] Adolphus, M. 2020. Information architecture and knowledge architecture. Emerald Group Publishing. Retrieved August 9, 2020 from http://www.emeraldgrouppublishing.com/librarians/info/viewpoints/info knowledge architecture.htm

[13] Meza, D. May 2016. Knowledge architecture: It's importance to an organization. NASA Johnson Space Center. Retrieved August 9, 2020 from https://appel.nasa.gov/wp-content/uploads/2016/06/Meza-David.pdf

[14] Hinton, A. 2014. What we make when we make information architecture. In *Reframing Information Architecture*. Springer, Switzerland, Chapter 8, 103–117. DOI: https://doi.org/10.1007/978-3-319-06492-5_8.

[15] Varaee, T., Habibi, J., and Mohaghar, A. 2015. Presenting an approach for conducting knowledge architecture within large-scale organizations. *PLoS ONE* 10, 5 (May 2015), e0127005. DOI: https://doi.org/10.1371/journal.pone.0127005.

[16] Kesh, S. and Ratnasingam, P. 2007. A knowledge architecture for it security. *Communications of the ACM* 50, 7 (July 2007), 103–108. DOI: https://doi.org/10.1145/1272516.1272521.

[17] Conway, M. 1967. Conway's law. Retrieved August 9, 2020 from http://www.melconway.com/Home/Conways Law.html

[18] Shrivastava, S. V. and Rathod, U. Categorization of risk factors for distributed agile projects. *Information and Software Technology* 58 (February 2015), 373–387. DOI: https://doi.org/10.1016/j.infsof.2014.07.007.

[19] Stankovic, D., Nikolic, V., Djordjevic, M., and Cao, D.-B. A survey study of critical success factors in Agile software projects in former Yugoslavia IT companies. *Journal of Systems and Software* 86, 6 (June 2013) 1663–1678. DOI: https://doi.org/10.1016/j.jss.2013.02.027.

[20] Wiggins, A. 2020. The twelve factor app. Retrieved August 9, 2020 from https://12factor.net/

[21] Li, Z., O'Brien, L., and Yang, Y. 2014. Impact of product complexity on actual effort in software developments: An empirical investigation. In *Proceedings of the 23rd Australasian Software Engineering Conference.* (Sydney, Australia, April 7-10 2014). ASWEC '14. IEEE Press, 170–179. DOI: https://doi.org/10.1109/ASWEC.2014.38.

[22] Jones, C. February 2014. Programming laws and reality: Do we know what we think we know? Retrieved August 9, 2020 from http://www.drdobbs.com/architecture-and-design/programming-laws-and-reality-do-we-know/240166164?queryText=language%2Blaw